# Auditing Digital Platforms for Discrimination in Economic Opportunity Advertising

SARA KINGSLEY, Carnegie Mellon University
CLARA WANG, Carnegie Mellon University
ALEXANDRA MIKHALENKO, Carnegie Mellon University
PROTEETI SINHA, Carnegie Mellon University
CHINMAY KULKARNI, Carnegie Mellon University

Digital platforms, including social networks, are major sources of economic information. Evidence suggests that digital platforms display different socioeconomic opportunities to demographic groups. Our work addresses this issue by presenting a methodology and software to audit digital platforms for bias and discrimination. To demonstrate, an audit of the Facebook platform and advertising network was conducted. Between October 2019 and May 2020, we collected 141,063 ads from the Facebook Ad Library API. Using machine learning classifiers, each ad was automatically labeled by the primary marketing category (housing, employment, credit, political, other). For each of the categories, we analyzed the distribution of the ad content by age group and gender. From the audit findings, we considered and present the limitations, needs, infrastructure and policies that would enable researchers to conduct more systematic audits in the future and advocate for why this work must be done. We also discuss how biased distributions impact what socioeconomic opportunities people have, especially when on digital platforms some demographic groups are disproportionately excluded from the population(s) that receive(s) content regulated by law.

## 1 INTRODUCTION

Social networks are a major source of information about hiring and economic opportunities[1] and impact the choices people have, and more generally, who gets what in society. The question is whether certain demographics in society are presented with opportunities that others are not. Digital platforms, including social networks, employ algorithms that decide, often in real time, what content is displayed to users based on data and assumptions about users and the relevance

---

[1]In 2015, Pew Research reported that "79 percent [of Americans had] utilized online resources in their most recent job search, and 34 percent said that these online resources were the most important tool available to them." See: https://www.pewresearch.org/internet/2015/11/19/searching-for-work-in-the-digital-era/

---

Authors thank Grace Bae, Sayan Chaudhry and Justine Cho, of Carnegie Mellon University, for their research assistance. A special thanks to Rediet Abebe, Julia Cambre, Caitie Lustig, Six Silberman, and Franzi Roesner, for their time and advice, and to the Mechanism Design for Social Good (MD4SG) community. To our anonymous collaborators, thank you for your dedication to pursuing questions that seek to ensure civil rights and equity in economic opportunity is a lived reality for everyone; hopefully, the end result of the long hard work is that our world is more just in the future.

---

of what is displayed in the content of the ad creative itself. As a result, answering the question of whether demographic groups are disproportionately shown some opportunities requires examining hundreds of thousands of algorithmic decisions; i.e. conducting platform-wide algorithmic audits.

Digital platform audits have been generally hard to conduct[23, 34, 35]. First, much of the data about content and users is available only to platform operators, not third-party auditors [31, 34, 35]. Second, even when some data is available, auditors need to augment this data with detailed categorical information of interest. For example, if a platform operator gives researchers access to all employment ads published on the digital platform, to conduct a meaningful audit, it is necessary to engineer features that provide information about the employment opportunities, i.e., what kind of job is being advertised. An auditor might not discover certain biases without this information, such as if platform shows ads for high-skilled (and high-paying) jobs to men, and low-skilled (and low-paying) to women. Third, and finally, audits must be replicable so findings can be verified by other researchers[35].

Our paper presents a methodology and software for auditing activity in advertising on digital platforms. Specifically, we build an auditing toolkit for the Facebook platform. Facebook currently provides a programmatic API to access ads.[2] Our audit leverages the Facebook API to get relevant ads-and-data about their distribution among user demographics. Between October 2019 and May 2020, we collected more than 141,063 advertisements.

For the audit, the API data was augmented with category labels, which were automatically inferred by classifiers we developed. These labels allow researchers to rapidly investigate ad content distribution along different categories of interest. Our classifiers, which we make available open-source, were used to classify ads by marketing categories that are regulated by law or policy (i.e., housing, employment, credit, political, and other). Once labeled, the demographic distributions of the advertisements were analyzed for bias. Our audit goes beyond previous experiments, and yields a statistical distribution of advertising content among users of social media by demographic.

***Policy implications***. We note that advertisers are prohibited from displaying credit or job opportunities disproportionately across demographic groups under current U.S. law (primarily through the Equal Credit Opportunity Act [18, 30] for credit opportunities, and the Civil Rights Act of 1964 [15] regulations for jobs). As a result, Facebook (and other social networks) ban targeting people by age and gender when advertisements are for housing, employment or credit (so-called HEC ads). However, our auditing results suggest that if advertisers published their ads as HEC advertisers, observed differences in the distribution of ads, if not a result of direct demographic targeting, then the bias is achieved through other means such as the platform optimizing the distribution (e.g. to maximize clicks.). It is also possible that advertisers did not mark themselves as HEC advertisers; or, otherwise managed to circumvent the platform's rules. Finally, it is also possible that Facebook only includes some ads selectively in its API, and other ads (which we were not allowed to access) corrected any biases in distribution.

In any case, an audit suggests several implications for future policy work. First, if advertisers circumvented platform rules, stricter checks on advertisers (including spot-checking ads) may be warranted. On the other hand, if advertisers were compliant, it would suggest that policies and targeting prohibitions are insufficient, and algorithms for ads distribution must be modified.

## 2   RELATED LITERATURE

This paper draws on prior work in three related areas: studies of algorithmic biases in advertising; methods for uncovering statistical biases using auditing in non-digital platforms; and legal judgments that suggest policy implications for digital platform audits.

---

[2]https://www.facebook.com/ads/library/

## 2.1  Bias in advertising are hard to measure

### Ad Explanations

Perhaps the simplest method for discovering if ads are disproportionately distributed to certain demographics is by analyzing "ad explanations" provided by platforms. Ad explanations inform users of **some but not all** of the reasons why they were shown a particular ad. Explanations also signal some information about biases in distribution, and have been used in legal cases as evidence. For example, in legal proceedings against Facebook, the Communication Workers of America [28] used screenshots of ad explanations as evidence that the platform allowed advertisers to target employment ads based on their age (Figure 1).

Ad explanations have limited ability in measuring distributional biases. Explanations tend not to disclose all the traits advertisers used to target people[5, 17]. In addition, Facebook users are not always aware that ads have links to explanations about why they were shown a given advertisement. When users are aware of these links in ads, they have reported being "confused by the ambiguity of the explanations" [17]. That is, ad explanations do not always make transparent to users the targeting methods or reasons why a platform sent them a particular ad.

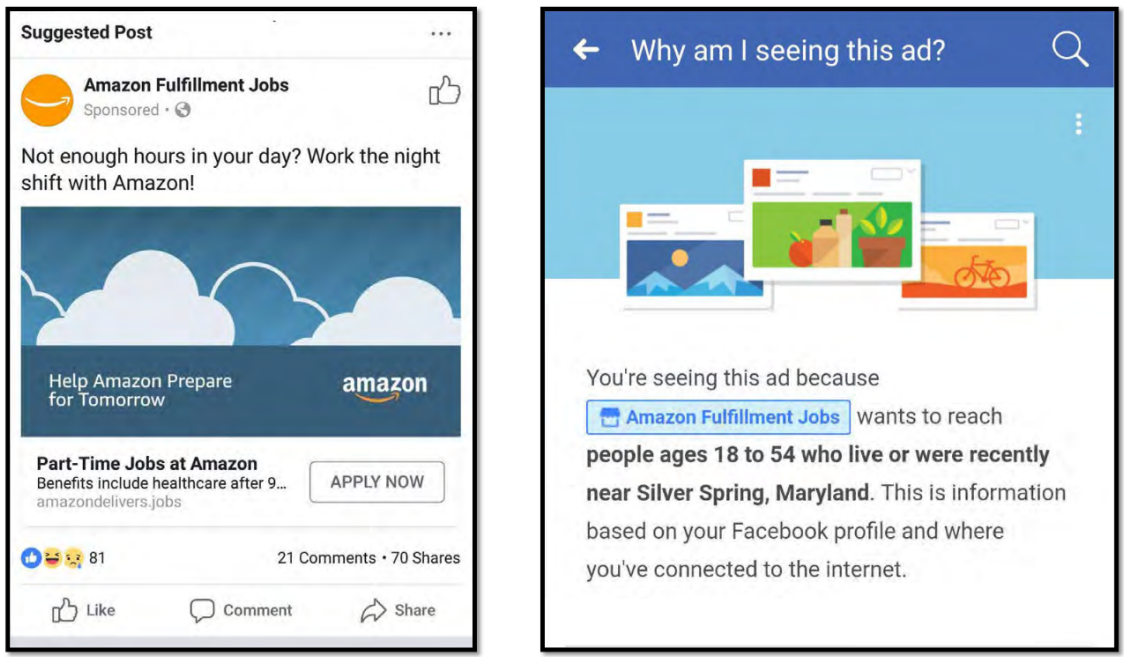

Fig. 1.  Employment Ad and Ad Explanation from Communication Workers of America Civil Rights Lawsuit against Facebook. Available online: https://www.onlineagediscrimination.com. Last accessed: Tuesday, June 30, 2020.

### Planned experiments

Planned experiments are a method that allow us to learn how advertisers and digital platforms could target users [4, 5, 21, 39]. In a planned experiment, researchers create ads with specifically controlled attributes. For example, researchers will create identical ads, but change one factor about a (treatment) ad while not changing that factor in the other (control) ad. After collecting data about the distribution of the ads, any differences observed will be used to infer how the changed factor in the (treatment) ad correlates with any differences from the control. Hence, by virtue of the research design, planned experiments can uncover the targeting methods that make it possible for advertisers to reach specific demographics, and the platform optimizations (that advertisers do not control) that result in disparate allocations of advertised economic opportunities. [4].

In Ali (2019) and Sapiezynski (2019), the researchers published experimental ads to Facebook, using the Look-alike Audience (LAL) tool. On the Facebook ad portal, the LAL tool allows advertisers to upload lists of and data about customers that have user traits that the advertiser wants to target [4]. After publishing the experimental ads, Ali et al. analyzed the distribution by gender and race (U.S. voter data was used to infer race). In their study, bias was observed. More importantly, the finding of bias indicated that the LAL tool could be used to target Facebook users by their demographic traits. In another planned experiment, Lambrecht and Tucker (2019) also published STEM employment ads on Facebook [21]. The design of the experimental ad content was varied to observe how that affected to which demographics Facebook would send the ads. Their study found that more men were shown ads for STEM employment opportunities. Older women were found to be particularly disadvantaged in the ad distribution. The finding suggests that digital platforms rely on stereotypes ("men prefer computing jobs more than women") and/or the biases of labor markets ("more men are employed in computing jobs") to decide how to optimize the distribution of content among users.

Ultimately, despite their utility in identifying problematic features (e.g., Look-alike audience tools, certain optimizations), planned experiments cannot describe the statistical distributions of the ads that a platform serves as a whole. This is because advertising experiments are typically only able to collect data about those to whom the experimental ads reached, and not data about the demographics that were shown ads published by other advertisers.

## 2.2 Discrimination: economics and legal theories

In this section, different theories of discrimination from economics and law are briefly described, and particular applications to digital contexts are discussed.

### Statistical discrimination.

In the economics literature, statistical discrimination describes the behavior of firms (or agents) when they base hiring and employment decisions on information they have about an entire demographic. The idea is that this discrimination is rational [19]; that is, firms or economic agents act on what they *believe or assume* about an entire group when making decisions where they have incomplete information about individuals. Economic theories of statistical discrimination suggest that decision-makers who statistically discriminate do not do so because of animus or an "intrinsic adversity to any particular group *per se*" but merely to improve the perceived quality of decisions [19]. While there is substantial evidence that human decision-makers do have habitual, implicit biases that do not improve the quality of the decision [12], this theory is helpful for thinking about the ways that algorithms distribute opportunities. Algorithms need not have an "animus" towards particular individuals, but may still lead to biased decisions, especially if inputs to computational systems that are biased themselves.

### Legally Protected Attributes

In US law, demographic traits that should not be used in making decisions about regulated economic opportunities are called "protected" attributes (cf. Table 1) [13, 14]. US law generally disallows for disparate treatment based on protected attributes for outcomes relating to employment, housing, credit, and other economic opportunities. Related to the current work, under US law, it is unlawful to discriminate against persons in marketing for certain economic opportunities (e.g., housing, employment, credit) on the basis of demographic attributes that are legally protected.

Table 1. Legally Protected Demographic Attributes in the US

| Attribute | |
|---|---|
| Age | Persons age 40 years and older |
| Sex or Gender | Includes persons who identify as non-binary and/or transgender |
| Sexual Orientation | Includes LGBPQ+ |
| Race | Includes persons who identify as belonging to more than one race |
| Color | Includes discrimination based on complexion |
| Disability | Includes physical, neurological medical and mental health identities |
| Genetic Information | Not limited to information known from genetic testing |
| Pregnancy status | Pregnancy is also a form of gender discrimination |
| Ethnicity | Includes persons who identify as Hispanic, Latinx, Jewish |
| National origin | Includes discrimination based on 1st spoken language |
| Religion | Includes discrimination against any religious group or belief |
| Political affiliation | Includes 3rd parties |
| Military or veteran status | Includes reserve members, ROTC |
| Citizen status | Includes persons who are not legally recognized citizens |
| Association | It is unlawful to discriminate against persons based on their association with a protected class |
| Witness or Whistleblower | It is unlawful to discriminate against persons because they were a victim or witness of discrimination and reported it |

**Disparate treatment**

Disparate treatment is a form of discrimination where decisions that consider or account for an individual's demographic attributes in ways that adversely affect their opportunities [22]. For example, if an employer only offers paid maternity leave to female employees, this adversely treats men who are employed at the company. Offering paid parental leave only to one gender, or only to persons who identify as male or female, is discriminatory; such a policy intentionally uses a legally protected demographic trait (gender identity) to determine eligibility for an employment benefit.

**Disparate impact**

Even policies and decisions that are facially neutral because they are uniformly applied to every demographic can still have different impacts on individuals from different demographic groups [22]. For example, an employer may deny paid parental leave to new parents, and apply this policy uniformly to all its employees. However, this policy could have a disparate impact on women, who may need to take more unpaid time off for unpaid parenting than men. Such policies are said to have disparate impact, and are illegal in the US when they discriminate against legally protected demographic classes (see Table 1).

**Proxy Discrimination**

Proxy targeting is when attributes of users that are statistically correlated with protected demographic traits are used to include or exclude them from the target population [33]. Proxy targeting can thus be used as an implicit method to make decisions about who will receive ads when legally protected demographic traits are not used explicitly by a platform's algorithm to make ad decisions.

In the United States, it is unlawful to use proxies for legally protected demographic attributes to discriminate against protected classes in marketing for certain economic opportunities (e.g., housing, employment, credit). For example, because US zip codes (postal codes) are so strongly correlated with race and ethnicity, it is unlawful to exclude persons from receiving ads for housing opportunities based on their zip code. Similarly, targeting "new college graduates", "fresh" or "young professionals" is an illegal proxy for age discrimination [3]. Proxy targeting need not be limited to demographic variables: requiring that applicants are able to "stand up for long periods of time", "walk" or be "quick on their feet", when recruiting for a desk job that does not require the use of legs or feet, could constitute discrimination on the basis of disability.

### 2.3   Auditing can uncover biases, improve policy enforcement

**Bias auditing in non-digital contexts** Audits have traditionally been used to uncover housing [7] and employment discrimination. Such audit studies use either paired-testing or correspondence audits, or resume study research designs[20].

In both paired-testing [7] and correspondence audits [8], near-identical candidates apply to jobs or housing; ideally, candidates are equal in every respect except in the targeted demographic trait (e.g. race or gender) [7, 20]. Data is then collected (such as whether candidates are offered a job or qualify for housing), which then yields evidence of discrimination, if any. Resume studies are similar, except that they create resumes of two fictitious candidates (materially identical in every respect, except the target demographic), and apply to the same jobs. Again data such as interview offers are used to evidence discrimination. Unlike paired testing, because resumes are fictitious, they can be used to test a wider variety of traits (even where a matched pair of individuals is hard to find).

These auditing methods have been used as legal evidence of discrimination in many employment and housing cases. Resume studies in particular can surface statistical discrimination. We posit that digital auditing methods such as ours can yield similar benefits in digital contexts.

### 2.4   Algorithm Audits

Artificial intelligence (AI) audits are referred to as "algorithm audits"[34] and are described as "the collection and analysis of outcomes from a fixed algorithm or defined model within a system"[34].

[1] **Input analysis** investigates whether the data used to train algorithms bias outcomes are biased or discriminatory. Studies in the medical literature, for instance, have been conducted with research participants who were all men. Despite this, the findings have been used to construct diagnostic criteria for entire populations. This has led to disparate outcomes in identifying heart disease and autism among women.

Using unrepresentative data to build models or decision-making criteria for entire populations is problematic. For example, researchers Joy Boulamwini and Timnit Gebru (2018) conducted an audit of facial recognition software [10]. In response, some of the investigated companies implied in public statements that the observed errors in classifying the faces of Black men and women were a result of using non-representative data to build the technology [34]. In alignment with what Boulamwini and Gebru have evidenced, as well as in other researchers in the machine learning community, it is worth stating that data is rarely the only source of bias in sociotechnical software applications, machines or systems.

[2] **Output analysis** investigates if the outcomes of a system are biased or discriminatory. First, researchers analyze if algorithms used in a system produce disproportionate impact among

demographics. An algorithm can be perfectly unbiased mathematically-speaking, and its inputs also, while the outputs affect populations differently.

*2.4.1   AI methods for audits.* A sock puppet audit replaces actors in traditional audits with scripts that pretend to be users or create fake traffic to a specific site [7, 38]. Using fake data, a sock puppet audit aims to test if an algorithm is biased. For example, one study found racial discrimination in Airbnb by using a sock puppet audit method where users that where identical in all aspects except name applied to various hosts [32]. The study discovered that those with names used more often by African-American individuals were rejected at a much higher rate than those with names used more by White Americans [32]. One downside of sock puppet auditing is the large amounts of data needed to demonstrate that a program or service is discriminatory. While this is made easier with the help of computer programs, it may cause problems later if researchers violate the Computer Fraud and Abuse Act or any Terms of Service of the platform [38].

## 2.5   Public policy and legal considerations

In this section, we outline the policy context against discrimination, and potential policy benefits of digital audits. To keep our discussion focused, we outline the policy context in the United States, but other countries and regions (such as the EU) have policies with similar intent.

*Legal protections against discrimination.* In the United States, protections against discrimination are guaranteed under the Civil Rights Act of 1964 [6], and follow-up legislation including the Fair Housing Act [29] and the Equal Credit Opportunity Act [11]. Some states have additional protections against discrimination. Overall, these laws prevent discrimination based on demographic traits, such as race, religion, sex, disability, national origin, and attributes such as genetic information.

These laws consider differences in *opportunity* to be acts of discrimination, just as differences in actual benefits provided to individuals. For instance, under the Equal Credit Opportunity Act, it is not necessary to refuse credit to people based on demographics to be considered discriminatory; instead it is sufficient for the creditor to have advertised to or discouraged people from applying for credit based on their protected demographic traits. Similar protections exist for employment and housing opportunities. As such, we consider advertisements to be a means for conveying opportunities; showing people fewer employment or credit ads in essence robs them of economic opportunity.

*Reactive legal frameworks.* Enforcement of anti-discriminatory regulation in the United States is largely reactive. For instance, job seekers must demonstrate that a difference in employment opportunities exists at a particular employer, rather than employers having to guarantee that their practices are non-discriminatory. The difficulty and cost of proving discrimination have made it difficult to systematically address discriminatory practices. However, if digital audits can be conducted automatically or at low cost, it may be possible to monitor practices efficiently, and ultimately, improve processes that work identify where systematic change is required.

## 2.6   Operationalizing legal protections on social networks

Civil Rights regulations imply special protections for opportunities in housing, employment, and credit (the "HEC" categories). Recent litigation has operationalized these protections on social networks, through a combination of technology and platform rules for ad targeting [2, 28].

In a March 2019 settlement, Facebook established "a separate advertising portal for creating housing, employment, and credit ("HEC") ads" that has "limited targeting options" for advertisers [2]. Through technological operationalization, Facebook removed options that permit advertisers to explicitly target users by gender, age, "multicultural affinity", and zip-code level or geographic targeting of less than 15 miles in radius. Facebook's operalization of the settlement agreement also modified the "Lookalike Audience tool" [2] The modification made it so advertisers cannot explicitly target demographics based on legally protected attributes. Finally, Facebook's operalization of the settlement agreement created new platform rules. The instituted rules state that advertisers must create and publish HEC ads on the HEC portal [2].

Table 2. Advertising Class Definitions

| Class | Definition |
|---|---|
| Housing | Advertises real estate property for rent or sale |
| | *Excludes mortgage, home financing or loans* |
| Employment | Recruits applicants for job openings |
| | *Includes job fairs, employment agent listings, employment services* |
| Credit | Advertises about credit, e.g., cards, rates, records; |
| | loans, e.g., auto, student, home, personal, business; |
| | insurance, e.g., life, medical, home, car, pet, disaster. |
| Political | Any ad by or about political candidates, campaigns, |
| | elected officials, elections, or policy/political agendas |
| Other | Any ad **not** defined by the other categories |

However, advertiser targeting is not the only source of disproportionate ad distribution – platform optimizations may cause biases too [4, 40]. Unlike ad targeting, rather than applying to a particular advertisement or advertiser, platform optimizations lead to statistical biases. As a result, they may not be prevented through rules limiting advertisers. Audits such as ours, along with established metrics, may help operationalize protections against such statistical biases.

## 3 SYSTEM DESCRIPTION

Our system has three components: an ad-querying and logging script that interacts with the Ad Library and downloads ads; a set of classifiers for adding meta-information to collected ads; and a database that allows researchers to query collected ads.

*Query script.* Our Python-based script queries and downloads ads from the Ad Library API. The appendix lists the fields and parameters used for querying the API. In our script, the parameter "search terms" was used to conduct keyword requests for data. Facebook's API currently seems to limit requests to around 2,000 ads per keyword[3]. In addition, the script also allows users to specify Facebook pages of advertisers of interest (such as Monster.com for job ads). Our script currently only requests data for ads displayed in the United States, but requests both active and inactive ads (i.e. both ads currently running on Facebook platforms and those no longer being delivered.) For each query, we save the exact text of the ads returned, URLs of any ad images (but not the images themselves), and the distribution of impressions by age, gender and geography. We also logged the search terms used in each data request in our database, along with the date of the request.

*Augmentation.* The Ad Library allows keyword searches for specific topics, but these searches are not always relevant for legal/policy applications. For instance, searching for "jobs" or "hiring" and only wanting to view ads recruiting applicants for employment, however, one discovers that they must wade through many political and others advertisements, such as depicted in Figure. For legal or policy utility, downloaded ads need additional information.

---

[3]According to Facebook, the official limit is 5,000 ads per API request. However, we found that the API often returns an error and asks developers to reduce the amount of data they are requesting if more than 2,000 ads are requested at a time.

We allow users to enrich ads with this additional information through *augmentation*. One useful augmentation system is a set of classifiers that automatically classifies each ad into policy-relevant categories.

We include a set of classifiers in our system, as described in Table 2. As detailed in Section 4, these classifiers allow for reasonably nuanced classification: if an advertisement is about employment policy, for example, the instance is classified political, because the ad pertains to a political agenda or policy. On the other hand, an ad for a job position in an elected representative's office is classified as an employment opportunity, not a political ad (Figure 2). This distinction will hopefully guide policy and legal workers when using our database, and help to identify the advertisements that are pertinent to their work. A different augmentation system we are currently developing labels objects in ad images, for instance, "dog", "woman", "construction equipment" etc.

Note that Facebook already has an internal system to identify, moderate, and remove (ad) content that violates policy or law. However, our method and software allows for an independent identification for external auditors, with standard performance metrics. Facebook has not released this data about its internal system. Furthermore, the goals of auditors are inherently different from business goals of Facebook – including social justice concerns such as hate crimes[4], discrimination[5], genocide,[6], sexism, misogyny, gender harassment, racism[7], racial harassment[8], xenophobia, anti-religious sentiment, ableism, ageism, anti-Semitism[9], white supremacy[10], and violence[11] In addition, whatever system Facebook uses to monitor/moderate advertising content, the system is not usable for external reviewers or auditors who want to track inorganic content (i.e., advertisements) on the platform for legal, policy or human rights reasons. Our software/methods will hopefully fill this gap.

*Database.* Our database includes numerous tables that contain the original raw data retrieved from the Facebook API, and tables that have the data enrichers included, as well as engineered features that augment the API data. For example, the database contains time-series tables that have information about when an advertisement was created, and also, how long ads were displayed to users on the Facebook platform. From these tables, users of our audit toolkit can analyze ads by certain policy or legal events, such as the date when the Facebook settlement agreement with civil rights organizations took effect. In addition, the database contains the development, training and test data-sets used to develop our classifiers. Researchers may access and use these data tables to verify and replicate our results. Finally, the database has tables which contain ads only for a specified HEC class (e.g., housing, employment, credit). Users may download and/or browse these tables to monitor and explore information about HEC ads published on Facebook platforms. For a full list of database column information, as well as tables that will be made available to our audit toolkit users, please refer to the appendix of this paper.

---

[4]Glorioso, Chris, Sola, Kristina and Stulberger, Evan (January 10, 2020). "I-Team: Anti-Semitic Trolls Impersonate Rabbis, Stoking Hate after Hasidic Attacks." NBC, New York: https://www.nbcnewyork.com/investigations/i-team-anti-semitic-trolls-impersonate-rabbis-stoking-hate-after-hasidic-attacks/2257562/

[5]Kofman, Ava, and Tobin, Ariana (December 13, 2019). "Facebook Ads Can Still Discriminate Against Women and Older Workers, Despite a Civil Rights Settlement." ProPublica: https://www.propublica.org/article/facebook-ads-can-still-discriminate-against-women-and-older-workers-despite-a-civil-rights-settlement

[6]Stevenson, Alexandra (November 6, 2018). "Facebook Admits it was used to incite violence in Myanmar." The New York Times, New York, New York: https://www.nytimes.com/2018/11/06/technology/myanmar-facebook.html

[7]Levin, Sam (September 12, 2016). "Facebook temporarily blocks Black Lives Matter activist after he posts racist email." The Guardian: https://www.theguardian.com/technology/2016/sep/12/facebook-blocks-shaun-king-black-lives-matter

[8]Cohen, Elizabeth (November 1, 2019). "She was called the n-word and given instructions to slit her wrists. What did Facebook do?" CNN: https://www.cnn.com/2019/11/01/health/facebook-harassment-eprise/index.html. Guynn, Jessica (August 3, 2017). "Facebook apologizes to black activist who was censored for calling out racism." USA Today: https://www.usatoday.com/story/tech/2017/08/03/facebook-ijeoma-oluo-hate-speech/537682001

[9]Nelson, Blake (January 3, 2020). "Gov. Murphy pushes Facebook to do more to fight anti-Semitism." https://www.nj.com/politics/2020/01/gov-murphy-pushes-facebook-to-do-more-to-fight-anti-semitism.html

[10]Newton, Casey (May 31, 2019). "How white supremacists evade Facebook bans." the Verge: https://www.theverge.com/interface/2019/5/31/18646525/facebook-white-supremacist-ban-evasion-proud-boys-name-change

[11]Cope, Sophia and Mackey, Aaron (August 7, 2019). "Second Circuit Rules That Section 230 Bars Civil Terrorism Claims Against Facebook." Electronic Frontier Foundation (EFF): https://www.eff.org/deeplinks/2019/08/second-circuit-rules-section-230-bars-civil-terrorism-claims-against-facebook. Authors' note: United States domestic, and not only international, terrorist events are a problem on Facebook. We note there is sometimes bias in the reporting of what constitutes a terrorist event for the United States generally, and ask that our readers not take this article as characteristic or definitional of all terrorist violence.

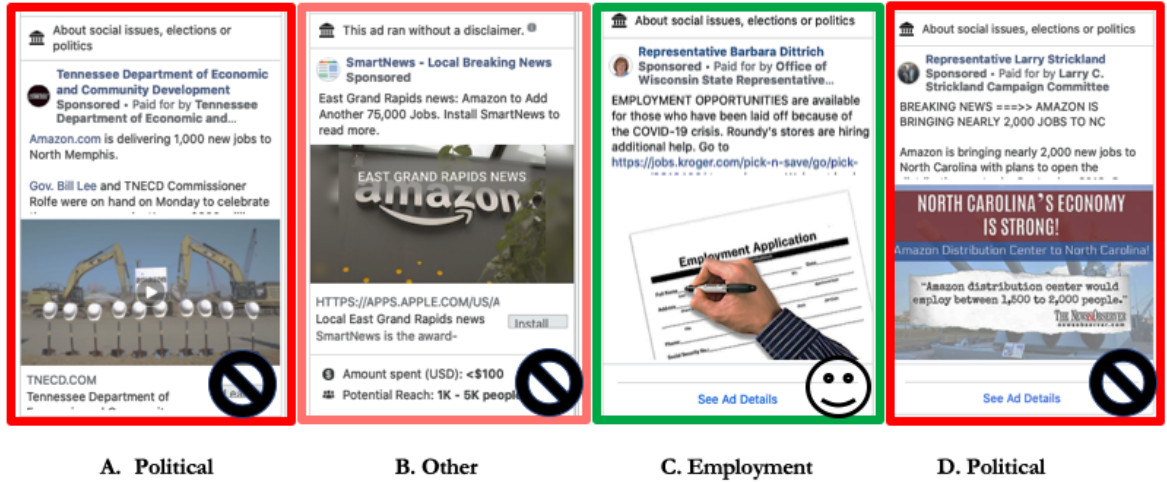

**Fig. 2.** Our classifiers categorize ads for policy/legal utility. An employment Ad (C) *recruits applicants for job openings at an employer.* In contrast, political and other ads (Images: A, C, D) sometimes inform users that an employer is creating jobs.

## 4 AUDITING METHODOLOGY

Below, we describe the specific audit we conducted using our system described above. For the audit, data was collected from Facebook's Ad Library API[12]. For each ad collected from the API, Facebook provided us with data about how the ad was distributed among demographic groups and by geographic region. Hence, the data allowed us to evaluate how the ads were distributed.

We augmented the data from the Ad Library by automatically classifying and defining ads by a primary marketing category (housing, employment, credit, political, and other). To train the models that were used to classify the ads, we manually added class labels to 3,767 ad campaign instances. The labels were defined in a way to describe an ad's intent or type of message. In addition to the primary classes, we built a logical ("Rules") algorithm to produce labels for ad sub-classes or types. For example, credit ads include sub-types or classes, such as for: student loans, debt relief, automobile loans, and home loans or mortgages. Defining sub-classes allows us to identify distributional differences of particular kinds of ads within each class, which might otherwise cancel each other out (e.g. some demographics may only see ads for debt relief, but not home loans.)

---

[12]https://www.facebook.com/ads/library

Table 3. Advertising Objective Options

| Objective | Facebook Description |
|---|---|
| Brand Awareness | Reach people more likely to pay attention to your ads and increase awareness for your brand. |
| Reach | Show your ad to the maximum number of people |
| Traffic | Send more people to a destination such as a website, app or Messenger conversation. |
| Engagement | Get more people to see and engage with your post or Page. |
| App Installs | end people to the app store where they can download your app |
| Video Views | Promote videos to raise awareness about your brand |
| Lead Generation | Collect lead information from people interested in your business |
| Messages | Get more people to have conversations with your business Messenger, WhatsApp or Instagram Direct |
| Conversions | Get people to take valuable actions on your website, app or in Messenger, such as adding payment info or making a purchase |
| Catalog Sales | Create ads that automatically show items from your catalog based on your target audience. |
| Store Traffic | create ads to generate traffic to your physical store location |

Table 4. Delivery Start Year, Number of Ad Campaigns

| Year | Ad Campaigns (#) |
|---|---|
| 2016 | 3 |
| 2017 | 26 |
| 2018 | 13,461 |
| 2019 | 52,418 |
| 2020 | 75,155 |

*Accounting for classification performance.* There are two sources of potential error in our audit. First, our classification models may have labeled ads incorrectly. Therefore, we evaluate the models performance in predicting advertising classes using multiple standard measures of goodness: accuracy, precision and recall. For most primary classes, the Naive Bayes models had precision and recall measures above 90 percent. For credit and employment ads, statistical measures of goodness neared 97-99 percent. Performance results held when our Naive Bayes models were tested on additional data-sets of derived labeled data (Table 7). These measures suggest that the classifiers adequately classify the collected ads to audit the distribution.

A second source of error may be that we may have omitted some relevant keywords and advertisers for querying the Ad Library. This source of error is extremely hard to measure as it deals in unknown unknowns. However, we tried to minimize these errors by expanding the set of keywords used over time, based on the copy of ads that we retrieved from the Ad Library. However, we did not notice any substantive changes to the overall statistical distributions we observed based on inclusion/exclusion of particular keywords, so this source of error is likely small.

### 4.0.1 Labelled Datasets for training classifiers.

Table 5. Advertising Class Distribution, Hand-annotated data from October to December 2019.

| Class | Percent | **Example Ad Text** |
|-------|---------|---------------------|
| Housing | 20.61% | "4 bed / 2 bath for $209,500" |
| Employment | 16.53% | "We have found the best job for you – Check it out FOOD SERVICE DIRECTOR, Huntingdon, PA" |
| Credit | 18.93% | "A planned visit to the vet? A quick trip to the eye doctor? Use the CareCredit credit card to finance your out-of-pocket health expense." |
| Political | 18.70% | "Health care just gets more expensive each year. Costs for the medicines we need keep going up. So, why did Rep. Chris Sprowls vote FOR a bill that will allow greedy insurance companies to charge huge medical bills to Floridians with pre-existing conditions?" |
| Other | 25.24 % | "Quit eating unhealthy foods to stretch your grocery dollars! SNAP helps millions of needy families get the benefits they need to purchase nutritious foods.Discover Discover if you qualify for the federal nutrition program and learn how to get started on your application today." |

*Manually Annotated Data-set #1.* From October 2019 through December 2019, we collected **25,228 distinct ad campaigns**, and **3,767 campaign instances** were manually classified. The labeled data was used to train a machine learning model to automatically label each instance in our database by advertising class (i.e., employment, credit, housing, political or other).

*Manually Annotated Data-set #2.* In January 2020, we collected an additional **16,805** ad campaigns from the Facebook Ad Library API. We hand-annotated labels for **11,584** advertising instances. The data was used to evaluate the Naive Bayes model trained on data from Oct. - Dec. 2019; and to train and evaluate a new model.

## 5 MODEL DESCRIPTION

We primarily use a naive Bayes model to classify ads according to whether the advertisements were for housing, employment, credit, political, or other opportunities. (The other opportunities class corresponds to "Uncategorized" ads as described by Facebook. However, these are better seen as opportunities other than housing, employment, credit or politics, so "other" seems to be more descriptive.) In addition, we use a rules-based model for subcategory classification.

Both models use the same features, namely the text in the body of the advertisement, expressed as a bag-of-words of unigrams and bigrams. These tokens are used as-is, and not stemmed or lemmatized.

**Naive Bayes**

The Naïve Bayes (NB) model uses a Bayesian model with prior probability of term's frequency, used smoothing, and a multinomial distribution. Ads were classified directly as one of credit, employment, housing, political, or other classes. Table 7 shows model performance.

Table 6. Advertising Class Distribution, Hand-annotated data from January 2020.

| Class | Percent | Example Ad Text |
|---|---|---|
| Housing | 1.61% | "Mark Markelz of William Raveis Real Estate just listed this 6bd, 3ba, 3,240 sqft home in Bridgeport, CT." |
| Employment | 7.33% | "Are you a high school student looking for a job? Premier Health is hiring high school students in various positions throughout Southwest Ohio. These are great opportunities to gain valuable health care experience and career insights. These jobs include: – Good pay – Flexible hours – Great people – Health care experience. Start on the path to success and a promising future with a job at Premier Health. To view open positions and apply, please visit https://php.referrals.selectminds.com/et/tcGA0GlB /page/high-school-job-opportunities-65 Please share this post or tag anyone that may be interested." |
| Credit | 31.88% | "ATTENTION HOMEOWNERS: It is your last chance to take advantage of the current federal tax credit for Solar (decreases at the end of 2019). You could be eligible to receive tier 1 solar panels from Momentum Solar for $0 out of pocket if your zip code qualifies. See if you qualify - http://bitly.com/PA$_c$ost" |
| Political | 46.38% | "Time is running out! Help us end the year strong AND make a qualifying donation to claim the OR Political Tax Credit!" |
| Other | 12.83% | "Cyber Security is one of the fastest growing industries in the country. The need for qualified experts is huge. Learn about the flexibility of our ONLINE program here!" |

**Rules Model**

Our Rules model is primarily used for sub-category classification. It uses a matching algorithm to determine the class-value of an instance. The Rules algorithm takes a vector of terms and searches for those terms in the main document or body text of an advertisement. If matching terms are found, the algorithm applies the classes label tied to that vector of terms, and otherwise, applies a label indicating that the instance does not belong to the class.

For each category, a classifier learns a "term-topic" vector that contains unigrams and bigrams that commonly occur in the text of ad instances belonging to a particular class. Ads are classified as belonging to the class if they contain any of the terms in this term topic vector, and as not belonging to the class otherwise. Note that this classifier may output more than one label for an ad as all rules are run in parallel.

Table 7. Performance on Test Data for Naive Bayes model for ads from 2019 and 2020

| Year | 2019 | | | 2020 | | |
|---|---|---|---|---|---|---|
| Class | Precision | Recall | F1 | Precision | Recall | F1 |
| credit | 0.991 | 1 | 0.9955 | 1 | 1 | 1 |
| employment | 1 | 0.9792 | 0.9895 | 0.9677 | 0.9783 | 0.973 |
| housing | 1 | 0.7686 | 0.8692 | 0.98 | 0.8167 | 0.8909 |
| political | 0.754 | 0.8716 | 0.8085 | 0.7731 | 0.8288 | 0.8 |
| other | 0.8938 | 0.9662 | 0.9286 | 0.875 | 0.9396 | 0.9061 |

A similar set of classifiers (with the same features) was also created for each category of ads. We describe its use below.

## 5.1 Final classification for analysis

The Naive Bayes classifier has generally high performance. However, for policy/legal audits, precision is more important than recall – audits would like to know if there is some statistical discrimination for ads that are clearly in a particular category, rather than try and include all ads that may be so classified. Therefore, for our reported analysis, we created an additional rules model for each category of ads (Like our sub-categorization model, this may produce multiple labels for an instance.)

Overall, this results in a stricter inclusion test for each category. For example, ads that are more about credit opportunities for housing (e.g. mortgages, home improvement loans) than housing opportunities more likely labeled as credit ads. Because the rules-model is only used as a filter for the final analysis, it only increases precision measures from Table 7 (possibly reducing recall.)

For our analysis of sub-classes of credit ads, while we only report findings here based on this stricter inclusion test, whether or not ads were filtered did not significantly change the findings. Please inquire with authors for findings based on analysis that: (a) only classified the ads using the Naive Bayes model, (b) only classified the ads using the Rules model, and (c) only used in the analysis if the Naive Bayes models and the rule-based model agreed on at least one classification (e.g. whether or not the ads was a "housing" ad).

## Ethics

To our knowledge, we have complied with Facebook's policies regarding the use of data from its Ad Library, and more generally, platform Terms of Service (TOS). Facebook permits authorized users of the Ad Library API to publish research about Facebook advertising. We did not intentionally collect any Facebook user Personal Identifiable Information (PII). However, if a Facebook user published ads on Facebook-owned platforms using a Facebook page that is named after their real name, that name data might be included in the Ad Library, and hence, our database.

## 6 FINDINGS

Our research database contains 141,063 Facebook ad campaigns for the period under study for the audit and 1,722,559 observations of more than 80 variables for those ads. These advertisements were collected between October 2019 and May 2020, and ran on Facebook between 2016 and May 2020 (see Table 4.)

*Systemic Gender Discrimination across Facebook Advertisements.* From our investigation, we discovered the design of the Facebook advertising portal could bake discrimination into advertising against persons who do not identify their gender to Facebook or who identify as non-binary and/or who identify a custom gender that is neither male or female.

The Facebook advertising portal only allows advertisers to target men, women or all. On the ad portal, if an advertiser specifies their advertisement is for housing, employment or credit, the only audience selection possible for gender is "all." Facebook defines "all" as men and women. We do not know whether "all" includes the gender group that Facebook calls "unknown" (in our paper we refer to this group as "custom" since it includes persons who identify as non-binary and/or specify a custom gender). Across every type of advertisement, persons who Facebook classified as having an "unknown" or custom gender receive few if any advertisements, and when shown ads, are on average between 0% and 1% of the total demographic.

### Credit Advertisements

In our database, for the time period under study, there are 67,181 observations of 6,385 ad campaigns classified as credit by the Naive Bayes model. A total of 274 advertisers published the ads on Facebook. Despite the number of ads campaigns (6,385), only 146 of the advertisements displayed unique text in the main text of the advertisement. This means that few advertisers published most of the ads. Out of the ads published, many were advertisements for the same thing, even though the ads delivered as distinct advertisements on Facebook (the ads also had individual budgets). Not every advertisement with the same text in the ad delivered to identical proportions of demographic groups.

In the credit ads, there were only 208 unique website links embedded in the ads; many more ads had identical embedded URL links. Embedded website links, if clicked on by a user, open a new browser tab that navigates to the website at the URL link. Among the ads labeled as credit, the embedded website links might direct the user to the website where they may apply for financing or loans, or other credit opportunities.

*Gender Distribution.* An estimated 57.9 percent of credit ads were sent to a greater percentage of men; whereas, 42.1 percent of credit ads were displayed to a greater percent of women. No credit ads were shown to a greater percentage of persons labeled as having a custom gender identity.

The distribution of credit ads among gender identity groups is notable. First, more women are shown ads on average, as a percent of the total demographic, in every ad class except credit. Second, more women than men use Facebook in the United States. Researchers have suggested that the size of a demographic on Facebook could explain the distribution of advertisements among those demographic groups. If this were true, since more women use Facebook in the United States, more women than men would receive advertisements. However, for credit ads in our database, the Facebook user-population of women in the United States does not explain the ad distribution by gender demographic. Facebook would need to show more credit ads to women for the size of the user demographic to explain the distribution of the ads; and this is not the case. Facebook, therefore, is not likely distributing credit ads based on the representation of a demographic on the platform. Instead, the platform and/or advertisers seem to target a specific demographic (men) to show credit advertisements.

What is remarkable, and worth noting again, is that for nearly every class but credit, more ads have distributions that are skewed toward women, meaning women are a greater percentage of the total demographic. In stark contrast, across advertisers and ad campaigns, a greater proportion of credit ads were distributed to a greater percentage of men. As observed for every ad class, the distribution of credit ads is never biased toward persons identified as having a custom gender identity (in other words, persons who do not identify as male or female on Facebook).

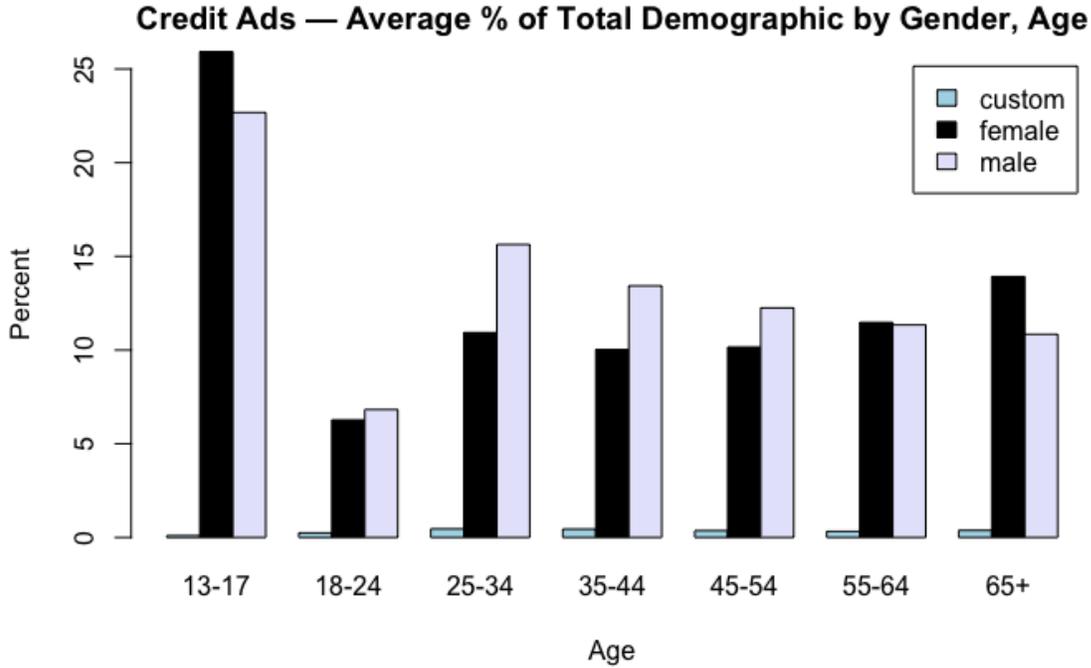

Fig. 3. Average Percentage Point of Total Demographic – Credit Ads by Gender and Age

Table 8. Credit Ads Only Shown to One Gender

|  | Men | Women | Custom gender |
|---|---|---|---|
| Ad Campaigns (#) | 209 | 210 | 2 |
| Advertisers | 29 | 32 | 2 |
| Funding entities | 23 | 22 | 2 |
| Embedded Websites (#) | 24 | 34 | 2 |

Table 9. In the above, are the number of campaigns that were only sent to: men, women or non-binary users (Facebook labels non-binary users, "unknown" gender). The number of advertisers and entities that funded the ads is shown below the campaign count. Finally, the number of embedded website links in the ads is given. Ads contain links to off-Facebook websites, such as to loan applications, and these links constitute embedded websites.

**Employment Advertisements**

In our database, for the time period under study, there were 165,853 observations of 13,250 ad campaigns that were classified as employment ads by the Naive Bayes model.

A total of 2,926 advertisers published the ads classified as employment in our database. Despite the number of ads campaigns, only 2,181 of the advertisements displayed unique text in the main text of the advertisement. There were 2,254 unique website links embedded in the ads. The embedded website links, if clicked on by a user, open a new browser tab that navigates to the website at the URL link. Among employment ads, the embedded website links might direct the user to the website where they may apply for the jobs advertised on Facebook, or other job opportunities.

Table 10. Proportion of Credit Ads by Age Skew

| Age Skew | Percent |
|----------|---------|
| 13 – 17  | .<1%    |
| 18 – 24  | 3.8%    |
| 25 – 34  | 30.8%   |
| 35 – 44  | 14.0%   |
| 45–54    | 13.5%   |
| 55–64    | 18.5%   |
| 65+      | 19.3%   |

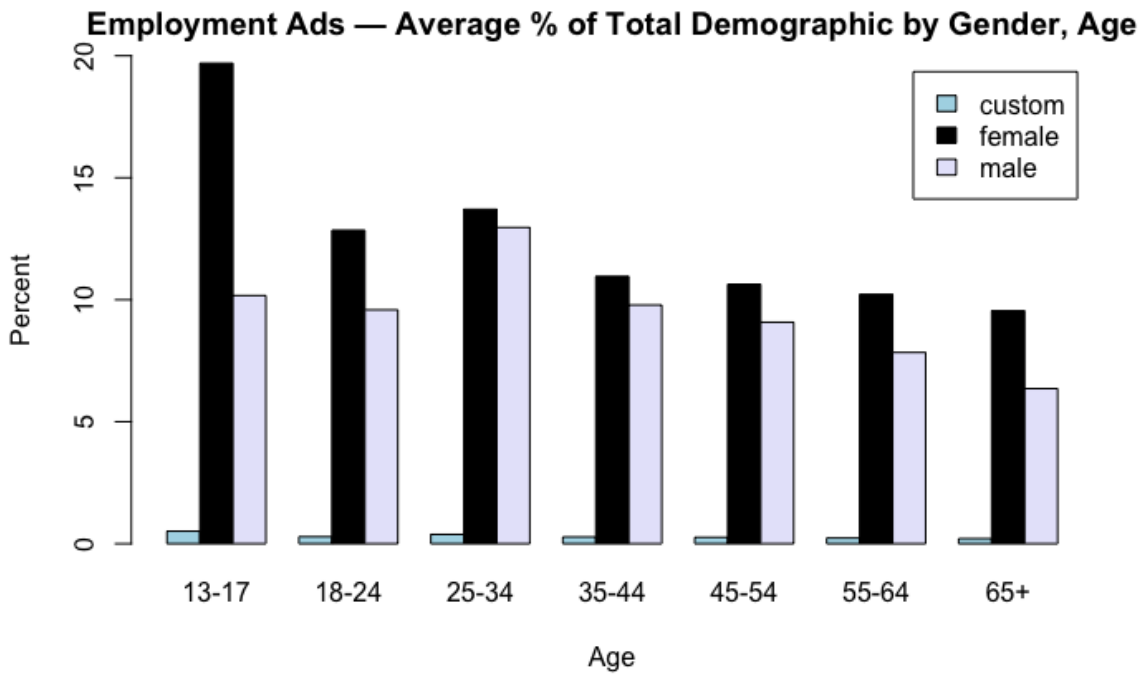

Fig. 4. Average Percentage Point of Total Demographic – Employment Ads by Gender and Age

*Gender Distribution.* An estimated 64.8 percent of employment ads were shown to a greater proportion of women (of the total demographic by gender group), while 35.2 percent were shown to a greater proportion of men. None of the employment ads were shown to a greater proportion of persons who do not identify their gender identity to Facebook and/or who do not identify as male or female.

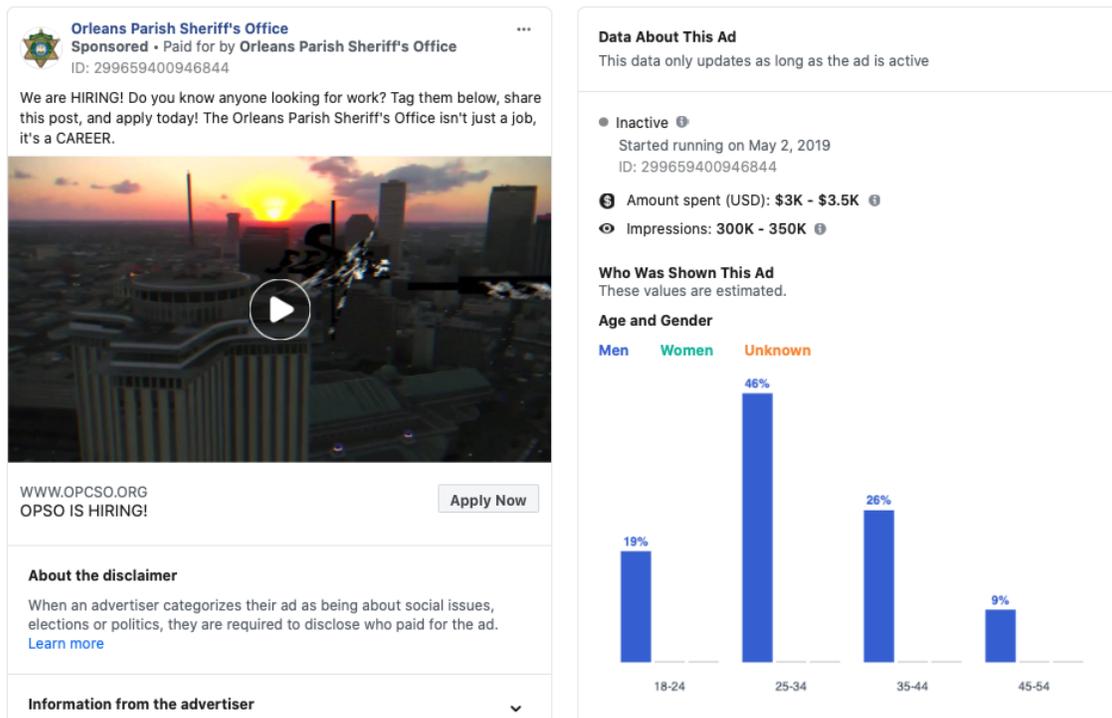

Fig. 5.  Employment Advertisement only sent to men on Facebook

Table 11.  Proportion of Employment Ads by Age Skew

| Age Skew | Percent |
|----------|---------|
| 13 - 17 | 1.2% |
| 18 - 24 | 16.3% |
| 25 - 34 | 30.4% |
| 35 - 44 | 15.5% |
| 45-54 | 11.8% |
| 55-64 | 11.6% |
| 65+ | 13.3% |

At least 49 ad campaigns for employment opportunities were displayed only to men and 58 were only shown to women[13]. Only 2 employment ads were displayed only to persons who were identified as having a custom gender identity.

---

[13]For the reported numbers about employment ads only sent to one gender, we based our estimate on instances that were classified as "employment" by both the Naive Bayes and Rules models. We did this because as reported in our Naive Bayes model performance metrics, there is some amount of error in classification. By using the agreement between both models, we capture a minimum estimate of employment ads that were only shown to one gender, and gain a little precision. In future work, we are working to fine-tune our algorithms further. Of note, using the Naive Bayes model, we report that more than 400 employment ads were shown only to men, more than 500 only to women, and only 3 to persons identified as having a custom gender

**Housing Advertisements**

In our database, for the time period under study, there were 22,140 observations of 1,820 ad campaigns that were classified as housing ads by the Naive Bayes model.

A total of 518 advertisers published the ads classified as employment in our database. Despite the number of ads campaigns, only 210 of the advertisements displayed unique text in the main text of the advertisement. There were 365 unique website links embedded in the ads. The embedded website links, if clicked on by a user, open a new browser tab that navigates to the website at the URL link. Among housing ads, the embedded website links might direct the user to the website where they may apply for an apartment or other housing opportunity advertised on Facebook, or other opportunities.

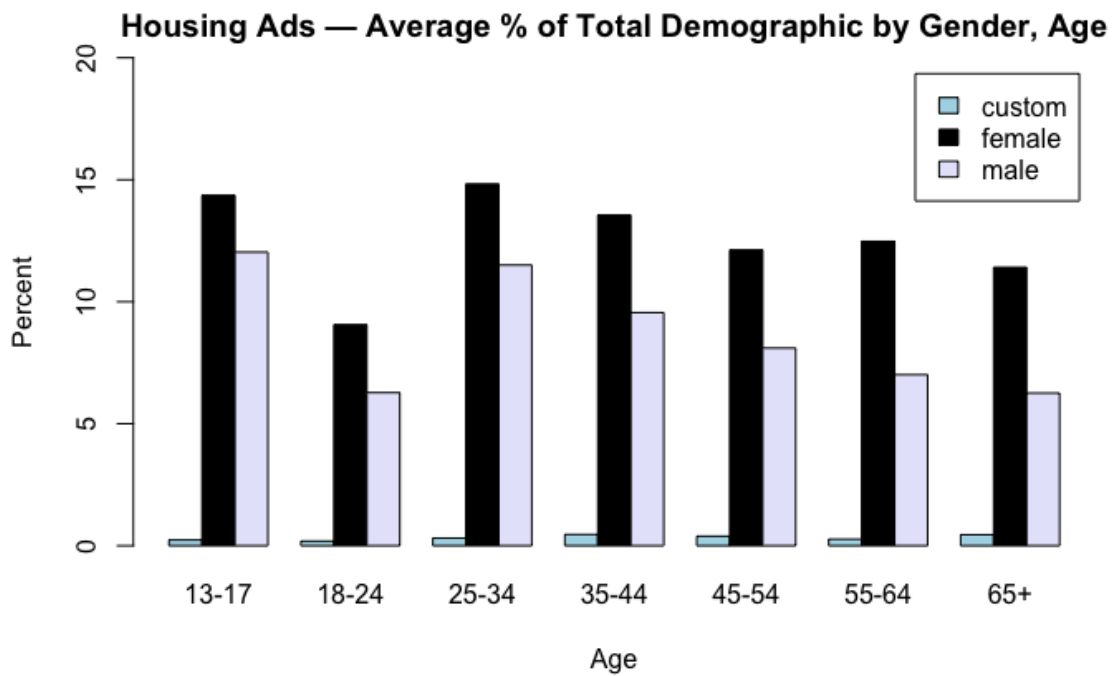

*6.0.1 Gender Distribution.* An estimated 73.5 percent of housing ads were shown to a greater proportion of women (compared to men or persons who identify as having no gender or a custom gender). Only around 26.5 percent of housing ads were shown to a greater proportion of men out of the total percentage of the demographic by gender, across all age groups. None of the housing ads were shown to more persons that Facebook identifies as unknown or custom gender (includes non-binary and/or trans).

*6.0.2 Age Distribution.* More housing ads (35.9%) were shown to a larger proportion of persons ages 25 to 34 years compared to housing ads in which other age groups comprised the largest proportion of the total demographic by age. An estimated 18.6% were shown to a greater percent of persons between the ages of 35-44, compared to the percent of the total demographic shown to other age groups; followed by persons ages 55 to 64, for which 14.5% ads favored this demographic. An estimated 9.6 percent of housing ads were sent to a greater percentage of persons aged 65 plus.

Table 12.  Proportion of Housing Ads by Age Skew

| Age Skew | Percent |
|:---:|:---:|
| 13 – 17 | 1.6% |
| 18 – 24 | 7.7% |
| 25 – 34 | 35.9% |
| 35 – 44 | 18.6% |
| 45–54 | 12.1% |
| 55–64 | 14.5% |
| 65+ | 9.6% |

Of all housing ads, the least were distributed in a biased fashion toward persons in the age ranges of 18 to 24 (7.7%) and 13 to 17 years (1.6%).

The trend in the distribution indicates that the age groups in which persons who attend college graduate and/or enter the labor market are favored to receive housing ads. Again, these numbers represent the percent of housing ads that have distributions biased or skewed in favor of an age group. However, if we examine the average percent of the total demographic that an age group typically represents in this distribution, the bias toward persons ages 25 to 44 remains.

**Limitations**

Our findings could be limited by our data collection method and the use of algorithms to classify ads by their primary and secondary advertising categories. In addition, we have not yet analyzed the data by dimensions available to us. For this paper, our audit work did not explore variance in the distribution of ads by geographic region or time period. We also did not analyze if the image or video media embedded in advertisements impacts the distribution of ads.

## 7   FUTURE WORK

There is a lot of potential in future work in this field. In the coming months, we are planning to study whether Facebook's inferences about the kind of ad content users prefer to be shown matches users' actual preferences. This is really important to study, as it is possible that Facebook's inferences on user preferences is statistically discriminatory and does not accurately represent the true preferences of the user in question. Facebook, in a lawsuit, suggests that it makes more sense from a business standpoint to show women ads for items such as cosmetics or clothing, while showing men ads for professional sports, as these are seen to be the general trend in preferences by gender [41, 43]. However, this is clearly a stereotype, and makes it harder for users to receive ads on actual areas of interest.

Furthermore, these comparisons on preferences are not really valid, as it might not be completely accurate to rank all forms of content in the same ranking. We could study whether it would be more representative of user preferences to rank preferences for seeing types of content that informs users about economic opportunities separately from a ranking about consumer opportunities. So, job opportunities would be ranked separately to consumer opportunities like buying clothes, cosmetics, sporting equipment and so on.

Another possible avenue of research could involve studying advertising trends and distributions in other countries around the world, to observe whether the labour market distribution is influenced by the distribution of ads about economic opportunities. For example, we could study advertisements in India, and observe whether the distribution of economic opportunity ads reflects facts about the labor market, such as the female labour force participation rate of 27% [36]. To accurately check this reflection, we could compare this statistic to the distribution of job opportunity ads on the platform in India.

## 8  DISCUSSION

Advertisements for economic opportunities are not distributed proportionally or equally on Facebook by gender and age among persons living in the United States.

**Gender**. Women are a larger percent of the total demographic shown advertisements. However, in evaluating the distribution of advertisements by class and sub-class, men are a larger percent of the demographic that receives ads for credit, particularly new lines of credit and financing. Meanwhile, ads for debt relief are disproportionately shown to women. Discrimination against women is endemic in credit markets. Research evidences that women are denied credit opportunities more often, and are offered worse terms for credit, if offered any credit at all. Importantly, in the United States, credit discrimination disproportionately impacts Black and African American women.

Persons who do not identify their gender identity to Facebook, or who identify as neither male or female, are rarely, if ever, shown credit ads of any type. When users either do not identify a gender on their Facebook profile, or identify as non-binary and/or trans, Facebook counts them as having an "unknown" gender identity. Across every class of advertisement, people identified as having an "unknown" gender, on average, are less than 1 percent of the total demographic shown any ad, if shown ads at all.

Discrimination against LGBPQ, transgender and/or non-binary persons is systematic and pervasive but hard to measure in the United States [9, 27]. Based on our audit findings, Facebook needs to: (a) publicly explain if the HEC advertising portals are designed to send ads to every gender by default, and (b) specifically, if the default options on the portal disproportionately exclude non-binary and/or persons who are transgender from receiving ads for economic opportunities. Facebook should consult with LGBTQ+ advocacy, policy and community organizations to discuss ways to measure and account for how ads for economic opportunities are distributed among persons who identify their gender as non-binary and/or transgender, gender identities specified by users, and those who choose not to identify. We recommend that Facebook consult the HCI Guidelines for Gender Equity and Inclusivity developed by Morgan Klaus Scheuerman and coauthors [26].

**Race and Ethnicity**. Facebook claims that their platform does not allow advertisers to target users by race or ethnicity, including when advertisements are for economic opportunities. At the same time, Facebook provides tools to advertisers to help them learn which user attributes are proxies for demographic traits, including race and ethnicity. Meanwhile, the Facebook Ad Library API does not provide researchers any information about the distribution of ads by race or ethnicity. Facebook claims the reason why the Ad Library does not provide this data is the privacy of users. However, the Ad Library API only provides aggregate and anonymous demographic data at the state and not town, city or county level. Therefore, it would be nearly impossible to identify any individual Facebook users from statistics describing how ads were distributed by race and ethnicity. If Facebook cares for the privacy of users, and specifically about the privacy of data about their race and ethnicity, Facebook should not provide advertisers tools that help them target users by race and ethnicity. Given Facebook's track record on civil rights, until then, we recommend that Facebook make aggregate and anonymous data publicly available via the Ad Library API about the distribution of advertisements by race and ethnicity at the state level.

Facebook allows advertisers to learn about and use proxy attributes to target users by race, but provides no information about the distribution of ads by race and ethnicity. Auditors, therefore, and the public, have no way to analyze if patterns of racial discrimination in credit markets are reflected in the distribution of economic opportunities on Facebook platforms.

**Responsibility for Harm**. In civil rights lawsuits, Facebook has argued that advertisers are to blame for discrimination on its platform, and that it merely provides "neutral" tools to advertisers. However, in a settlement of legal charges, Facebook promised to, and subsequently changed the design of the advertising portal.

Advertisers must now use the HEC advertising portal when publishing ads for housing, employment, and credit (so-called HEC ads). The HEC portal does not allow advertisers to target users by gender and age. The HEC portal also limits the proxy attributes that advertisers can use to target users. If changes to the advertising portal indeed prevented advertisers from using protected demographic or proxy attributes to target users, then we would expect that no HEC ads would be distributed only to one gender or age demographic group.

Our audit reveals that ads are not only disproportionately sent to one gender identity, but the distribution of ads is skewed by age, and importantly, some ads are still only sent to one gender. That ads for economic opportunities are still only sent to one demographic, while others are excluded, suggests that the changes to the Facebook advertising portal do not prevent discrimination in the distribution of HEC ads.

One possible explanation of this is that advertisers did not disclose that their ads were HEC related, and so were not subject to limitations of targeting. Second, it is possible the ad-distribution algorithm improperly optimized who should see advertisements, leading to systematic biases. Finally, it is possible that advertisers still managed to find proxy variables that allowed discrimination (see below).

Regardless of the specific cause, our research suggests that much of this bias can be reduced. We discuss some potential mitigation approaches below.

**Proxy Targeting.** Even if advertisers cannot target specific genders, races and ethnicities, the Facebook Audience Insights page allows advertisers to find proxies for these various attributes by displaying data like device activity, work, and education that can be filtered by gender, age, location, and multicultural affinity. While this may be a useful tool for analyzing and predicting the impact of a given ad, it also introduces the possibility of targeting not constrained by law. Thus, Facebook allows advertisers to reinforce current biases and inequalities by identifying proxies for various groups of people through Audience Insights data.

The quantitative impact of proxy targeting on the demographic distribution of ad viewers may be investigated more deeply in the future to determine if these proxies are a workaround for the regulations on discrimination and targeting. More investigation into ways to restrict proxy targeting should also be considered.

**New questions and opportunities** Our work also suggests new questions for fairness and user rights. We hope, by writing this paper, to start a discussion about when disproportionate distribution is irrelevant, and when it is harmful. For instance, is a 1 percent difference between the percentage of the demographics that were shown content advertising reasonable? What about 15 to 20 percent difference? In the remaining sections, we discuss considerations of harm, fairness and (in)equity in allocating resources.

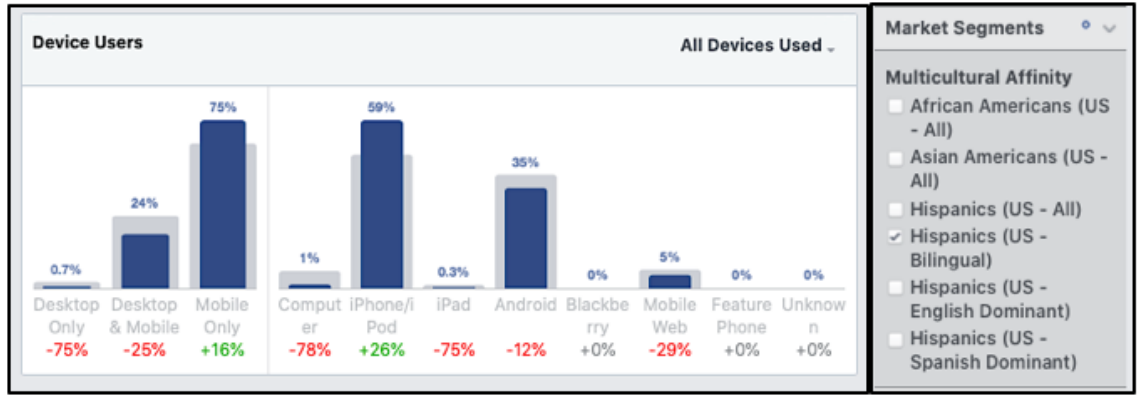

**Image 1:** Facebook's *Audience Insights Tool* enables advertisers to select behaviors, activities, demographic and other attributes and learn how much more likely a user audience is to have the attribute or preference. For example, **Image 1** shows that bilingual Hispanic Facebook users in the United States are 75 percent less likely to only use Desktop devices. Information like this is usable as a proxy variable in HEC advertisement targeting. Facebook *Audience Insights tool* is available on the web and was accessed for this paper as recently as May 30, 2020. See: https://www.facebook.com/ads/audience-insights

*Stereotypes in Algorithmic decision-making.* Algorithms are used in the decision making process in various different scenarios and contexts, including but not limited to: providing credit, hiring processes, housing loans, and more generally, advertisement distributions. Algorithms predict the future behavior of individuals using imperfect information/data that they have from past behavior of other individuals who belong to the same socio-cultural group [25]. If we take the example of credit advertisements, then we can see that this statistical discrimination perpetuates a reinforcing cycle over iterations. If the qualification of a certain group, for example women, was historically lower than a dominant group, then the algorithm will not preference individuals from the disadvantaged group in the distribution of credit ads among demographics. Further, as algorithms are written, developed and tested by humans, it is possible for personal biases and personally held stereotypes to seep in.

The distribution of advertisements for economic opportunities on Facebook reflect social stereotypes perpetuated in the United States. For example, advertisements for new credit lines are shown more to men, while advertisements for debt relief are shown more to women. Different types of job ads, which we briefly explored in this paper, are also distributed with an algorithm that shows signs of stereotyping. While women on the whole received more job ads than men, the positions advertised were traditionally roles filled by women like "secretary" or "nurse" whereas men were more likely to receive ads for traditionally masculine work such as construction. Thus, the division of labor by gender reinforces the general stereotypes of each gender.

*Legal and Ethical responsibilities of Platforms.* In accordance with Section 230 of the Communications Decency Act, "No provider or user of an interactive computer service shall be treated as the publisher or speaker of any information provided by another information content provider"[14]. In other words, a platform cannot in theory be held accountable for the content which is published on it – be it individual user posts or large-scale advertisement campaigns. However, since these platforms are the ones providing the tools for advertisements to be distributed, it is our belief that they should be responsible for designing these tools such that they cannot be misused.

---

[14] 47 U.S.C. § 230, url = "https://www.law.cornell.edu/uscode/text/47/230"

Although Facebook limits the direct targeting options for HEC ads, it still relies on advertisers to self-disclose that their ad falls into one of these categories, thus allowing an opportunity for exploitation. Proxy targeting also offers a workaround for some of the targeting limitations on protected classes. It can be argued that, since such actions are outwardly prohibited by the platform, Facebook should not be held accountable when advertisers choose to disobey its rules. However, our findings and the above discussion show that such misuses are not uncommon, and thus demonstrate structural flaws in the way Facebook handles advertisements on its platform. Indeed, the company has been previously involved in lawsuits on the premise of unfair advertisement distribution despite not being the entity which publishes the ads [24, 42].

The unfair distribution is important to consider, particularly when it comes to protected classes such as gender or race, because it may lead to reinforcement of existing societal inequalities and biases. If one ad for a STEM position is disproportionately targeted towards men, it will likely result in a man being hired for this position; if a hundred of different ads are, it might then provide further support for the stereotype of women not being interested in STEM jobs, and the reality of them not getting equal opportunities to apply for such positions. Because of this, we believe that it is important to hold large platforms accountable for the way they allow their often-massive user bases to be reached by marketing campaigns. Advertisers may choose to post ads which are dishonestly classified or targeted beyond the permitted categories using proxies, but these exploits are available solely because of the way Facebook functions as an advertising platform. This makes it Facebook's responsibility to maintain and update the tools it offers to advertisers if it wants to lower the availability of such exploits.

Ultimately, our software and method for auditing Facebook advertising supports and also contributes new directions to the algorithm and digital platform auditing literature. First, analyzing the demographic distribution of economic opportunity advertisements across advertisers, advertisement campaigns and over time is important. Preliminary findings from our Facebook audit indicate that evidence produced from advertising experiments do not necessarily generalize to describe overall trends in the distribution of advertisements on platforms. In other words, while experiments are useful for learning how features of digital platforms and ads could bias distributions on platforms, the studies do not necessarily inform us about how ads are distributed in the wild among demographic groups. Collecting and analyzing data provided by digital platforms can allow us to compliment experimental methods with data about the demographic distribution of ads. We applaud Facebook for making this data available, suggest that Facebook assess and make improvements to the Ad Library API, and encourage digital platforms that distribute or impact the economic opportunities people have, to create and release data APIs for transparency, accountability and auditing of their platform.

## 9   CONCLUSION

Over the course of history, legalized and illegal discrimination has segregated markets, institutions and communities [16, 20, 37]. At the heart of concern is that digital platforms and their advertising networks increasingly decide how to allocate economic opportunities among demographics, and that the historical nature of societal inequity is reflected in the outcomes. Disparities in the distribution of economic opportunity in the United States is a historic problem, disproportionately impacting persons who identify as Black and African American, Latino/a/x, LGBQ+, and disabled. Gender and age discrimination compounds the impact of historic and present-day discrimination. Digital platforms that also discriminate compound the impact of prior biases in society over time, and have also called into question the legitimacy of judiciary or legal systems, the possibility of criminal and civil justice, and the democratic process itself.

Remarkably, (digital) discrimination is not *always* viewed as overt or hostile. It sometimes manifests as an (algorithmic) preference for a familiar or favored group(s). At the same time, discrimination is also sometimes overt.[15] For example, our audit uncovered systemic bias in the allocation of opportunities on a major advertising platform, Facebook. In particular, persons classified as having a custom gender, meaning they either did not wish to disclose their gender identity to Facebook, and/or do not identify as male or female, were shown few if any advertisements for housing, employment or credit (so-called HEC ads). Women were also disadvantaged. Most credit ads were distributed to a greater percentage of men. Finally, the distribution of most HEC ads evidenced an age bias. These audit findings are problematic and indicate that digital platforms and advertising networks reproduce historical and present-day societal biases that marginalize demographic groups. Future research should investigate the extent to which our audit findings generalize to platforms and advertising networks where regulated economic opportunities are advertised.

Measuring harm in the distribution of economic opportunities that are advertised on digital platforms is hard. First, access to platform and particularly demographic distribution data is limited. Second, while legal rules establish causes for action under the law, U.S. federal rules tend toward a minimum standard, and ignore undesirable bias on digital platforms. Third, beyond legal minimums, we should also care about equity in the allocation of opportunities. However, it is challenging to measure and determine if a distribution is acceptable. People, communities, as well as societies, have different definitions of what is a permissible allocation of economic opportunities in society. For example, research shows that while "most White Americans accept basic principles of equal opportunity, at the same time, [they] resist the implementation of policies that would increase equality directly"[20].

Bias in advertising on digital platforms could be addressed by removing interface features that allow advertisers to target users by demographic traits. However, it is debatable whether removing such features would suffice to redistribute the balance of economic opportunity ads. Instead, digital platforms and advertising networks might need to take a proactive human-centered approach. For example, digital platforms could add features that allow users to instruct the platform to send them advertisements for economic opportunities only or most often when opportunities are more urgently needed, such as when unemployed or searching for work. Compare this to the status quo, at present, digital platforms permit advertisers to send job opportunities only to the already employed.

In researching if the allocation of advertising is a social problem, we adopted and emphasize the sentiment that "roles for computing in social change" include "diagnostic" work [1]. That is, research which uncovers undesired outcomes but forgoes technological solution-ism. In this role, the aim of computing work is to produce evidence for broader efforts that seek to create socially just systems. In this paper, we demonstrated a methodology to document and monitor how a digital platform allocates economic opportunities. Our findings indicated that digital platforms cannot simply, as they have done, tell advertisers not to use demographic targeting if their ads are for housing, employment or credit. Instead, advertising must actively monitored. In addition, platform operators must implement mechanisms that actually prevent advertisers from violating norms and policies in the first place. Governments also have a role in improving the status quo. Incentives for making platform and advertising network data available to third-party auditors are needed. Penalties under the law that discourage audits by criminalizing certain digital audit methods (e.g. web scraping) need to be removed. Finally, government regulators could ease business concerns by issuing regulatory and legal guidance, and by providing technical assistance to help platform operators comply with established rules.

## A   APPENDIX

### A.1   Description of Advertisement Database

Table 12, on the following page, lists the variables or features that are available for advertisements in our database.

Table 13. Variables in Ad Database

| Variable | Description |
| --- | --- |
| archiveID | ID that locates the ad in the Ad Library |
| ad_creation_time | Time and date (UTC) that the ad was created |
| text | Text displayed in the main body of the ad |
| url_caption | If an ad has an embedded website link, this is the text that appears in the link |
| url_description | If an ad has an embedded website link, this is the text the appears with the link |
| url_title | If an ad has an embedded website link, this is the text the appears as the link Title |
| ad_delivery_start_time | Time and date (UTC) an ad started running |
| ad_delivery_stop_time | Time and date (UTC) an ad stopped running |
| embedded_url | A permanent URL link to the location of the ad in the Ad Library |
| currency | Currency used to pay for the ad |
| funding_entity | Name of the person or organization that paid for the ad |
| impressions | Minimum & Maximum impressions |
| potential_reach | Minimum and maximum potential audience size |
| page_id | ID for the Facebook page that ran the ad |
| page_name | Name of the Facebook page that ran the ad |
| publisher_platforms | List of platforms that the ad was displayed on |
| region_distribution | Geographic distribution by U.S. state. Given as a percentage |
| spend | Minimum & Maximum money spent on an ad |
| age | age range for the observations in the instance/database row |
| gender | gender identity for the observations in the instance/database row |
| percentage_demographic | percentage of the total demographic shown an ad |
| predicted_label | Predicted marketing class of the ad |
| ad_subClass | Predicted subclass of the ad marketing class |
| max_percentage | Maximum percentage of the demographic sent an ad |
| max_genderDemographic | Gender that is greatest percentage of demographic |
| max_ageDemographic | Age group that is the greatest percentage of demographic |
| startTimeDayOfWeek | Day of the week that the ad started running |
| stopTimeDayOfWeek | Day of the week that the ad stopped running |
| adStartWeek | Week of the year that the ad started running |
| adStopWeek | Week of the year that the ad stopped running |
| adDuration_Seconds | Number of seconds that ad ran on the platform |
| adDuration_Minutes | Number of minutes that ad ran on the platform |
| adDuration_Hours | Number of hours that ad ran on the platform |
| adDuration_Days | Number of days that ad ran on the platform |
| adDuration_Weeks | Number of weeks that ad ran on the platform |
| adDuration_Months | Number of months that ad ran on the platform |
| adStart_Semester | Whether ad started running in the 1st or 2nd half of the year |
| adStop_Semester | Whether ad stopped running in the 1st or 2nd half of the year |
| startQuarter | Financial quarter that ad started running on the platform |
| stopQuarter | Financial quarter thatad stopped running on the platform |